\documentclass[twocolumn]{webofc}
\usepackage[varg]{txfonts}   
\usepackage{hyperref}
\usepackage{url}
\hypersetup{colorlinks=true,citecolor=blue,urlcolor=blue,linkcolor=blue}
\newcommand{\cherenkov}{\v{C}herenkov }
\newcommand{\cm}{\mathrm{cm}}
\newcommand{\mm}{\mathrm{mm}}
\newcommand{\um}{\mathrm{\mu m}}

\begin{document}

\title{Dual readout with capillary tubes – status and prospects}

\author{\firstname{Nicol\`o} \lastname{Valle}\inst{1,2}\fnsep\thanks{\email{nicolo.valle@cern.ch}} 
}

\institute{INFN, Pavia (IT)
\and
           on behalf of the HiDRa collaboration
          }

\abstract{
This paper discusses the ongoing development and testing of dual-readout fiber-based calorimeter demonstrators. An overview of the existing prototype will be presented, along with details on the assembly and integration strategies of the High-Resolution Highly-Granular Dual-Readout Demonstrator (HiDRa). A concise yet informative summary of simulation results and a perspective on upcoming beam tests will be discussed.
}
\maketitle

\section{Introduction}

The advancement of high-energy physics experiments, particularly after the LHC era, relies on the development of sophisticated detectors capable of precise measurements. Proposed future projects such as the Future Circular Collider (FCC) and the Circular Electron-Positron Collider (CEPC) aim to exploit the benefits of electron-positron collisions for high-precision studies of electroweak, Higgs and flavor physics. A critical aspect of these studies is the accurate measurement of hadronic jets which requires advanced calorimetry techniques targeting resolution as good as $30\% / \sqrt{E}$.

The dual-readout method has emerged as a promising technique for improving hadron energy reconstruction and has already demonstrated reliability and advancements over the last decades \cite{DREAM,ART}. This method involves the simultaneous measurement of signals from scintillating and \cherenkov light produced in the shower development, thereby correcting for the electromagnetic fraction of hadron showers on an event-by-event basis. Further improvements have been reached in the latest R\&D developments with the the integration of compact photosensors (Silicon Photomultipliers, SiPMs). This permits fine segmentation, enabling sophisticated event reconstruction through particle-flow and machine learning algorithms.

Different dual-readout calorimetry projects are currently under development; in this paper the  technology exploiting fiber light collection, with fibers embedded in capillary tubes, will be presented. The concept behind the design of a scalable and cost-affordable solution will be discussed, focusing on the activities ongoing to build a demonstrator with a large hadronic-shower containment and the expected performance in a real scenario.

\section{The electromagnetic-shower prototype}

\begin{figure}[h]
\centering
\includegraphics[width=6.5cm,clip]{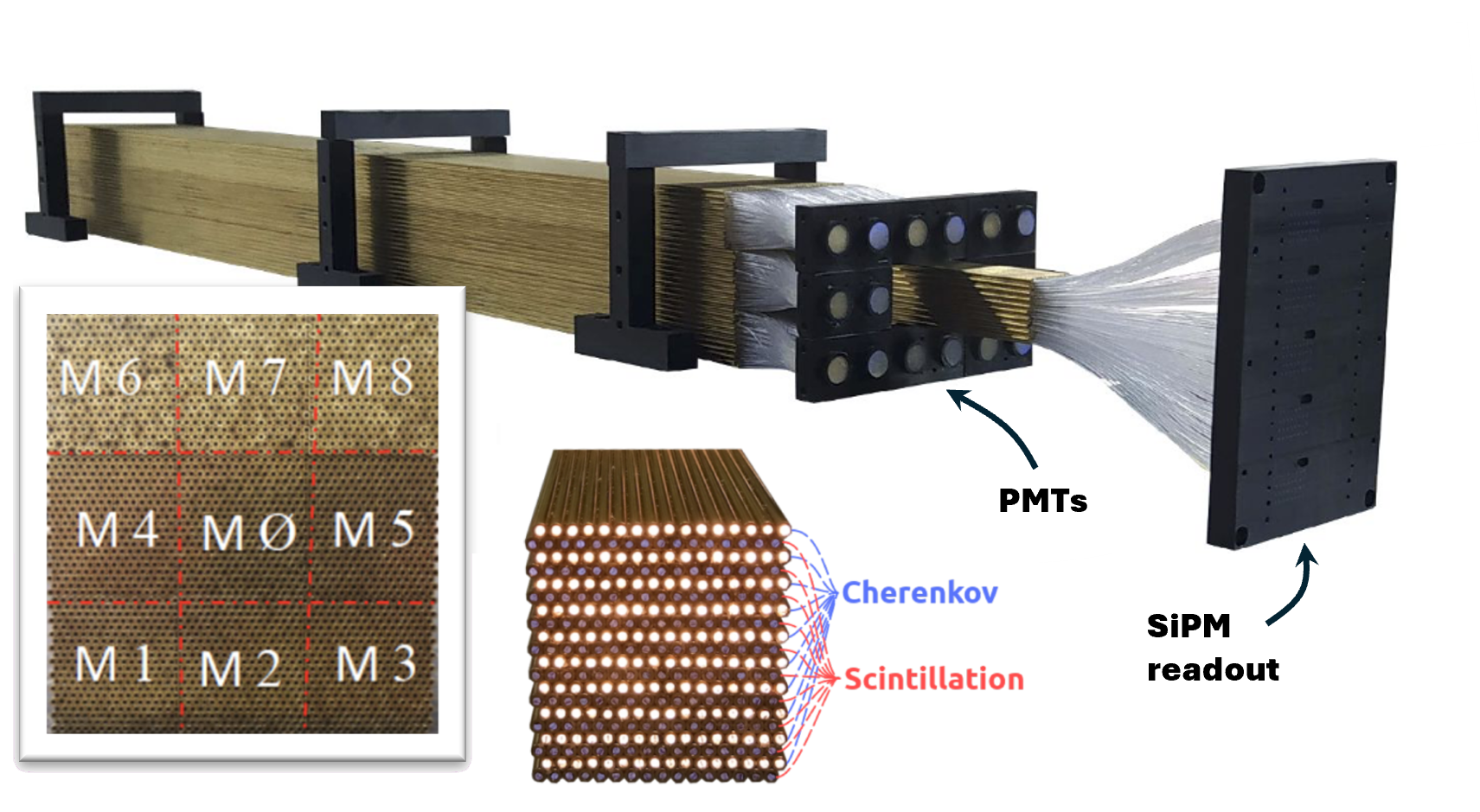}
\caption{Layout of the brass capillaries demonstrator assembled in 2021. The layout of the alternating rows of clear and scintillating fibers loaded in each of the 9 modules can be appreciated in the middle inset. }
\label{fig::EMpicture}       
\end{figure}

In 2021, the first demonstrator designed to validate the feasibility of single fiber readout with SiPMs in a dual-readout calorimeter was assembled. This demonstrator utilizes brass capillary tubes (63\% Cu, 37\% Zn), arranged in a honeycomb geometry and hosting alternating rows of doped (scintillating) and clear optical plastic fibers, which are responsible for collecting \cherenkov light. The capillaries have an outer diameter of $2\, \mm$ and an inner diameter of $1.1\, \mm$, resulting in approximately $63\%$ absorber material by volume, including the gaps between capillaries. The prototype, designed to contain electron-induced showers, features a frontal surface area of $100\, \cm^2$ and a length of $100\ \cm$. It employs a modular geometry, with the capillaries organized into 9 towers, each containing 320 tubes. The layout of the detector, its modules, and the fibers are illustrated in Figure \ref{fig::EMpicture}.

The readout approach features a hybrid solution with both Photomultiplier Tubes (PMTs) and SiPMs. Due to the dimensional constraints of the SiPM package at the time of construction, fibers from the central tower are routed to an interface board at the back of the calorimeter, used for an easier connection to the front-end boards. 
Each fiber of the central module, both scintillating and \cherenkov, is individually connected to a Hamamatsu S14160-1315PS SiPM, with a photosensitive area of $1.3\times 1.3\, \mm^2$ and  $15\, \um$ pitch, ensuring a proper dynamic range \cite{ROMUALDO}.
The peripheral modules, labeled M1 to M8 in the figure, are each equipped with two PMTs. Each of them collects photons from all scintillating and \cherenkov fibers within the module, respectively. A yellow filter is placed at the end of the scintillating fibers to filter out shorter-wavelength components of the spectrum, in order to mitigate the effect of the attenuation length within the plastic material and reduce the signal's dependency on the starting point of the shower.

\subsection{Testing on the beam}

The prototype has been tested with positron beams at the DESY laboratories in Germany in 2021, and then at the CERN SPS in 2021 and 2023. Highlights from the 2021 testing campaign are summarized in the following, while all the details can be found in \cite{TB2021}.

\begin{figure}[h]
\centering
\includegraphics[width=\columnwidth,clip]{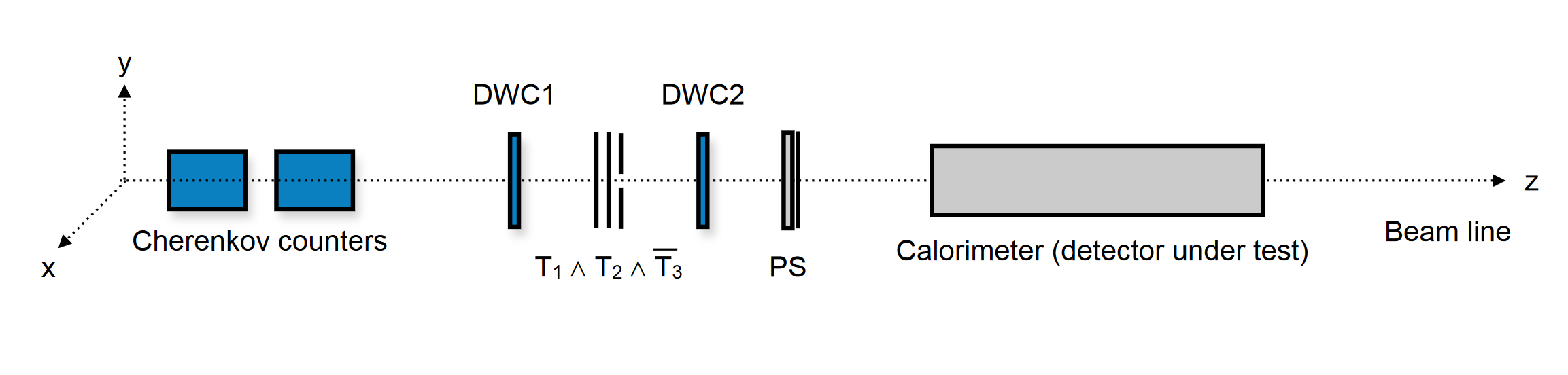}
\caption{Beam line setup for 2021 measurements at CERN SPS (not-to-scale). Taken from \cite{TB2021}.}
\label{fig::21beamline}      
\end{figure}

The test beam setup is schematized in Figure \ref{fig::21beamline}. It includes two \cherenkov counters for positron selection, Delay Wire Chambers (DWCs) to estimate the impact point on the calorimeter, scintillator plates for triggering and muon selection, and finally a lead-scintillator preshower detector (PS) for the  identification of high-energy positrons.

\begin{figure}[h]
\centering
\includegraphics[width=0.9\columnwidth,clip]{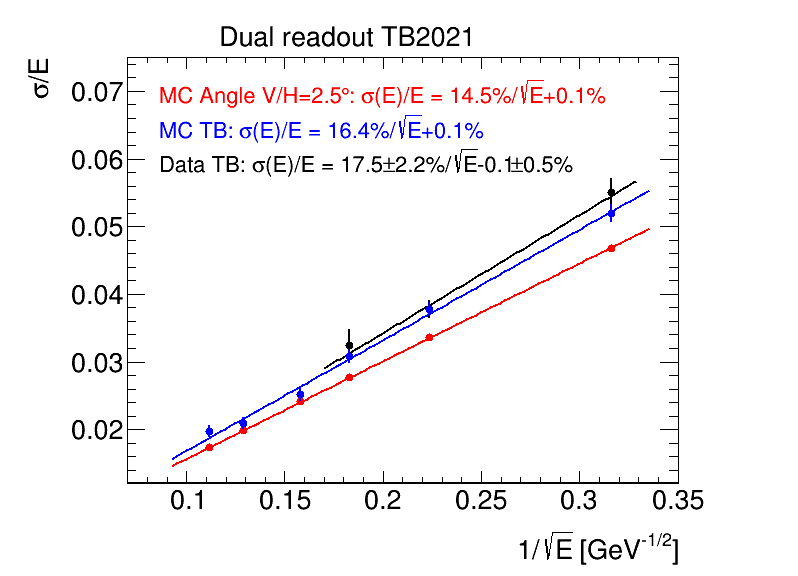}
\caption{Measured energy resolution. Data points (black) are available only for $E<30$ GeV, and are compared with simulation including the full beam setup (blue). Red points correspond to data simulated by tilting the calorimeter by an angle of $2.5^\circ$ (see text). Taken from \cite{TB2021}.}
\label{fig::21reso}      
\end{figure}

The calorimeter exhibited a good response, reconstructing the true beam energy within $1\%$ across the entire energy range of $10-120$ GeV. However, energy resolution estimates were limited up to 30 GeV, due to shower containment issues caused by the placement of the PS detector. Installation constraints within the experimental area meant that the PS could only be positioned 285 cm from the face of the calorimeter, resulting in an increased leakage and a worsened resolution.
At lower energies, a sufficient hadron rejection could be achieved using the upstream \cherenkov counters, permitting the removal of the PS. The calorimeter's resolution, evaluated as the ratio of the RMS width to the mean value of a Gaussian fit to the energy measurement distribution, is illustrated in Figure \ref{fig::21reso} for both experimental data and simulation.

In ref \cite{TB2021} a detailed study can be found on the dependence of the calibrated signal from the impact point position of the beam. A periodic modulation in the $y$ coordinate has been identified, corresponding to the spacing between different fibers, with a smaller amplitude observed in the \cherenkov signal due to the correlation between the direction of the emitted light and the electron's trajectory. 
Such a modulation is influenced by the number of fibers sampling the shower, therefore by the angle of tilt between the fibers and the beam line. A tilt angle of $2.5^\circ$ was found to effectively mitigate this effect. For this reason, the resolution measured in simulations at this tilt angle is also presented in Figure \ref{fig::21reso}, demonstrating a reduction in the stochastic term from $(17.5\pm 2.2)\%/\sqrt{E}$, measured on the test beam, to $14.5\%/\sqrt{E}$.

\begin{figure}[h]
\centering
\includegraphics[width=6cm,clip]{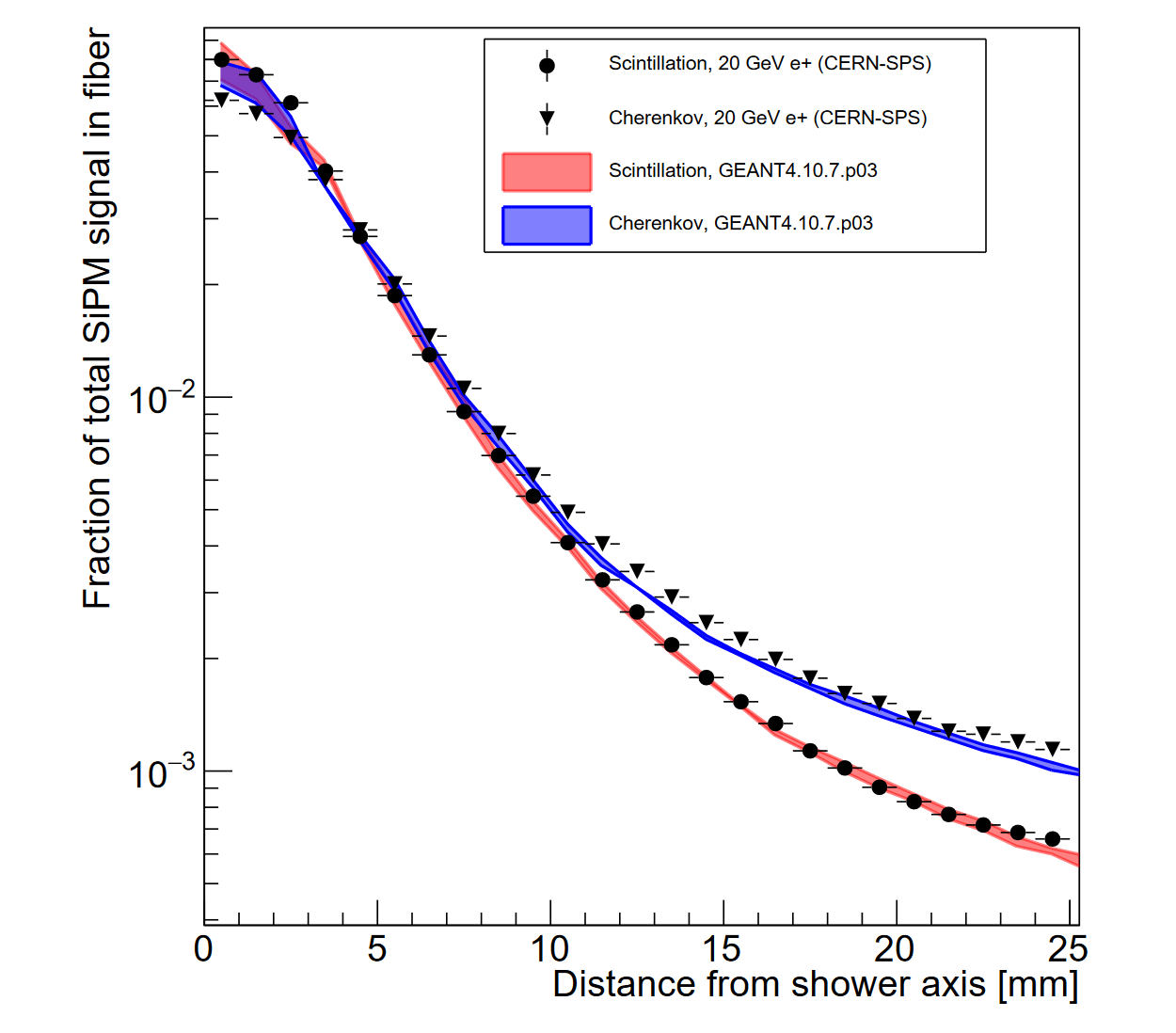}
\caption{Fraction of energy deposited in a fiber as a function of distance from the shower axis (20 GeV positrons). The broader profile of the Cherenkov signal is explained by the correlation between the emitted light and the direction of the charged particle, resulting in a dependence of photon collection within the fiber's numerical aperture on the stage of shower development \cite{TB2021}.}
\label{fig::21profile}      
\end{figure}

Despite the challenges and experimental constraints, key measurements were achieved, including a comprehensive validation of the Geant4 simulation used to reproduce the calorimeter geometry and the entire beam setup. Figure \ref{fig::21profile} shows the lateral profile of showers produced by 20 GeV positrons in the central tower. 
The implementation of single fiber readout with SiPMs enabled an unprecedented millimeter-level sampling of the shower development.  This high degree of accuracy in reproducing the shower profile at such a fine resolution scale is a further proof of the effectiveness of Geant4 simulations in guiding the design and optimization of future detectors.

Building on the knowledge acquired in the first testing campaign, the same detector was brought again to the SPS beam in 2023, with an improved setup (the preshower placed at $\sim 15\, \cm$ from the front face) and improved beam purity. The calorimeter was exposed to electrons, muons and pions in order to characterize the response to the shower core and validate the Geant4 hadronic models also in hadron beams.  The analysis of the collected data is being finalized and a significant improvement in the energy resolution for electromagnetic showers is expected.

\section{HiDRa}

\begin{figure}[h]
\centering
\includegraphics[width=\columnwidth,clip]{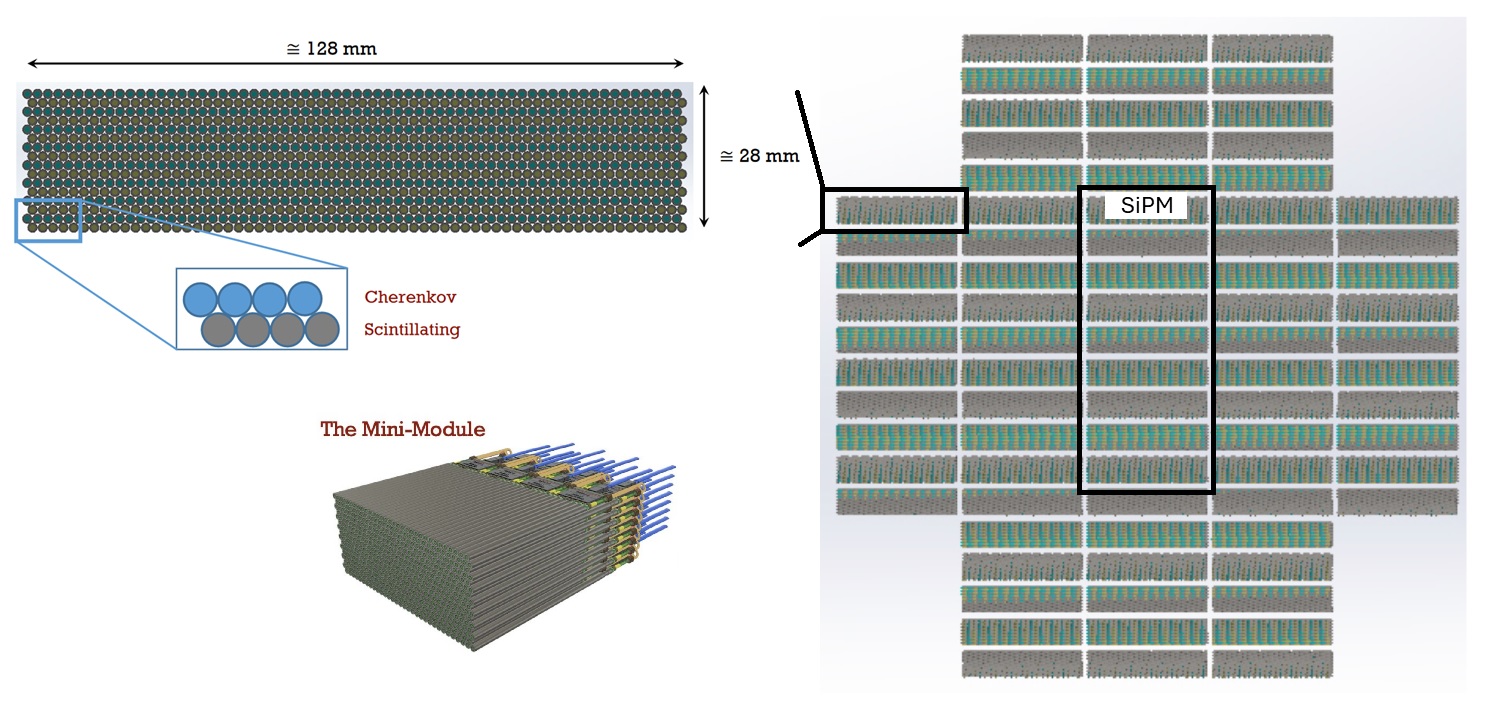}
\caption{Layout of the front of the HiDRa prototype. The elementary module made of $64\times 16$ capillaries with alternating rows of different fibers is displayed on the top-left. The bottom inset shows an example of the board system designed for SiPM readout.}
\label{fig::hidralayout}      
\end{figure}

HiDRa (the High-Resolution Highly-Granular Dual-Readout Demonstrator) is a prototype designed to contain hadronic showers, currently being assembled at the INFN laboratories in Pavia, Italy, in collaboration with various INFN groups. The construction approach closely follows the previously described methods, yet featuring differences in scale and choice of the materials.
This modular prototype will be composed of 80 independent modules, made by a matrix of $64 \times 16$ stainless steel capillary tubes and alternating rows of scintillating and clear (\cherenkov) fibers. The total frontal area of the calorimeter will measure $65 \times 65\, \cm^2$, and it will extend to a length of 250~cm, equivalent to more than ten radiation lengths.  As for the smaller prototype, the tubes have inner and outer diameters of $1.1$ and $2\, \mm$, respectively. Illustrations of the layout can be found in Figure \ref{fig::hidralayout}.

The ten innermost modules will be equipped with SiPM readout, while two PMTs will be used for each of the outer-shell modules. They will collect photons separately from each type of fiber within their respective modules. Specific models of SiPMs have been selected for scintillation and \cherenkov signals to optimize performance. S16676-10(ES1) Hamamatsu SiPMs, $10\, \um$ pitch, have been chosen for scintillating fibers, allowing for a larger dynamic range. S16676-15(ES1), $15\, \um$ pitch, will be used with the clear fibers to improve the photodection efficiency where the light yield is significantly lower. Peripheral modules employ Hamamatsu PMTs (R8900 for scintillating channels and R8900-100 for \cherenkov channels).

A total of 10240 SiPMs will be needed. To reduce the readout system complexity, an analog sum of eight adjacent channels will be implemented directly on the front-end boards. This will result in 64 independent channels for both scintillating and \cherenkov fibers per module.  The CAEN A5202 FERS system, exploiting two WeeROC Citiroc 1A ASICs, is used as readout board. Two FERS boards will operate each single module.

\subsection{Mechanical assembly and integration}

The assembly of HiDRa started in November 2023. At the time of writing (July 2024), 36 modules have been completed and assembled, forming a more compact prototype equipped with PMT readout only. It will be tested at a beam line in 2024 for preliminary operation tests and Geant4 simulation validation.

\begin{figure}[h]
\centering
\includegraphics[width=\columnwidth,clip]{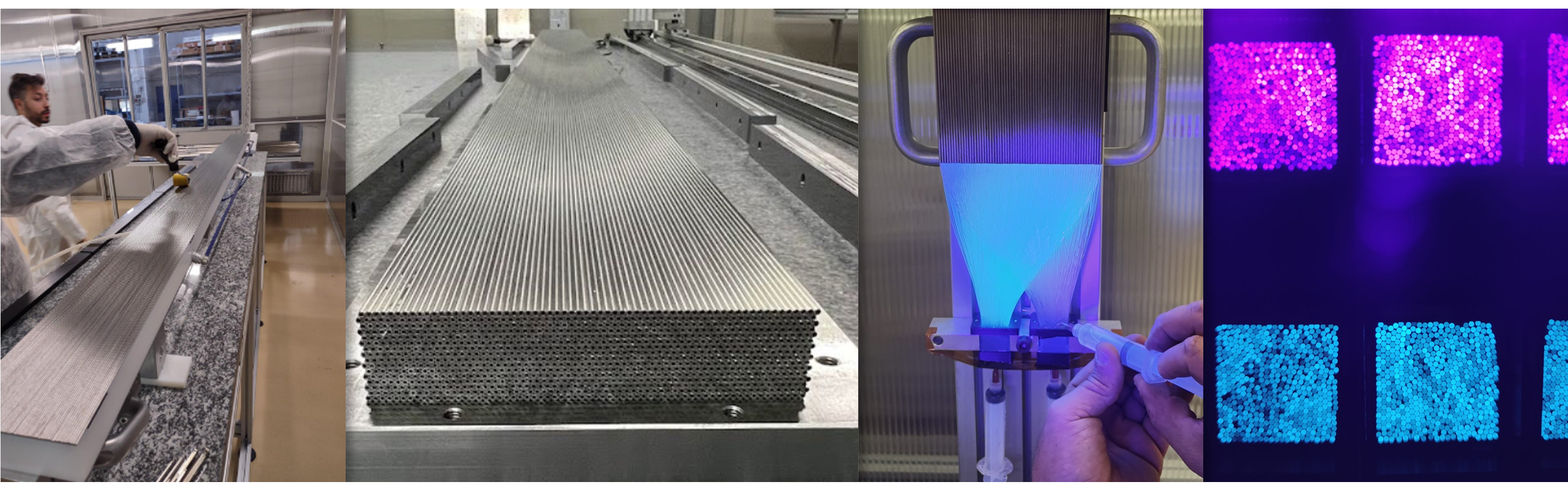}
\caption{HiDRa assembly site at Pavia INFN laboratories. From left to right: one layer of the module being glued; one HiDRa module ready for fiber loading; fibers grouped in bundles; detail on the glued scintillating (violet) and \cherenkov (blue) fibers, ready to be coupled to PMTs.}
\label{fig::assembly}      
\end{figure}

In addition to aiming for excellent physics performance, considerable effort has been devoted to study and refine a scalable construction method. This must permit an efficient integration of mechanical components within compact volumes, which is crucial for the development of future full-coverage calorimeters.
A comprehensive set of mechanical tools and quality assurance methodologies has been developed to facilitate the assembly of the prototype. The tubes are handled using a vacuum system, aligned within a reference tool and glued with a bi-component Araldite 2011. Each layer, consisting of 64 tubes coated with a layer of glue, is carefully aligned on top of the previous one, building up to a total of 16 layers that together constitute a module. The final assembly step involves loading optical fibers. Once loaded, the fibers are grouped in bundles of 512 into a plastic support, they are glued and polished before being coupled with PMTs using optical grease to enhance signal transmission. Pictures of the assembly site are shown in Figure \ref{fig::assembly}.

\begin{figure}[h]
\centering
\includegraphics[width=0.9\columnwidth,clip]{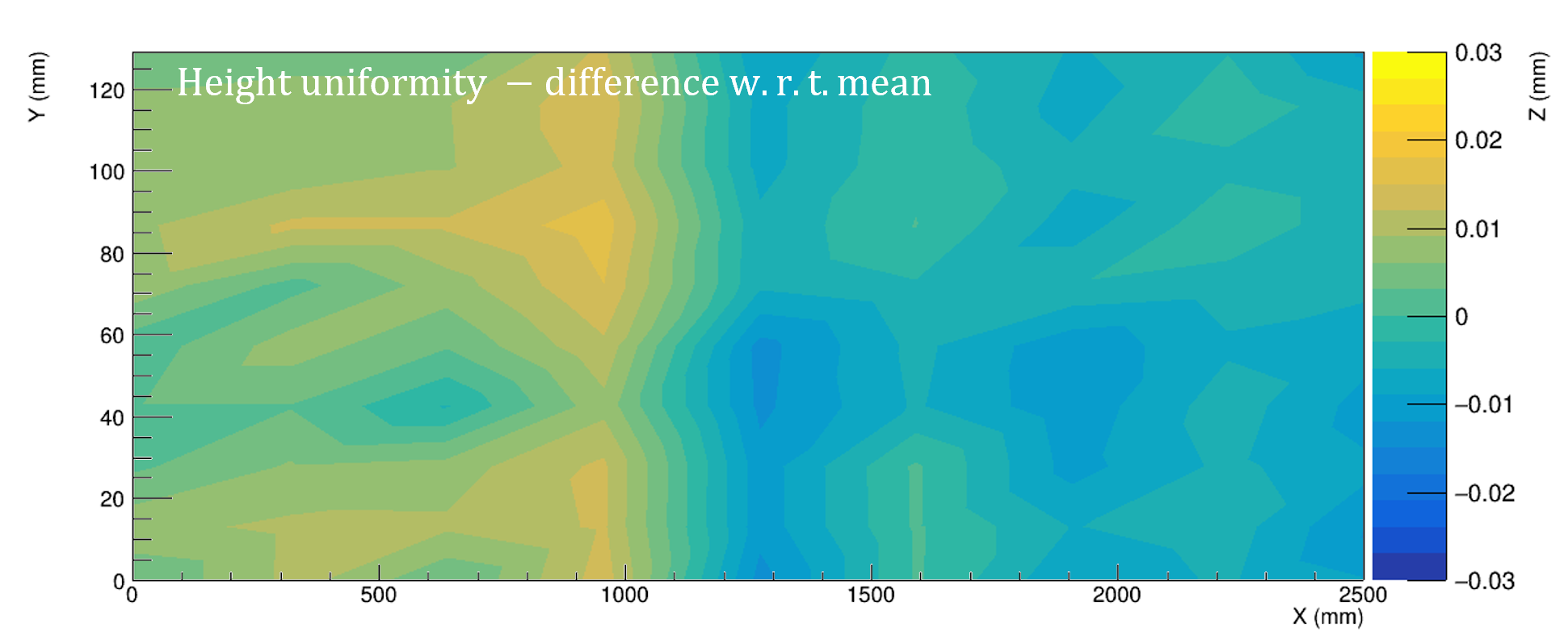}
\caption{Typical planarity distribution of one HiDRa module across its surface, after the assembly. The slight and tolerable non-uniformity in the longitudinal direction can be explained by assuming that different pressures were applied to the module after gluing.}
\label{fig::qcplanarity}      
\end{figure}

Quality control is an essential aspect of the process. Among various checks, an automated tool scans the module's surface to assess its planarity. Figure \ref{fig::qcplanarity} illustrates the typical distribution of thickness across the module's surface. The variance among the sampled points can be as low as $10\, \um$, comfortably within the acceptable limits for the final assembly. The average height of each module is $28.2\, \mm$.

Parallel to this, significant advancements have been achieved in the design, study, and qualification of SiPMs and their integration into the system. While waiting for the delivery of the final components, the integration of the front-end boards is being thoroughly tested using 3D-printed prototypes to ensure flawless functionality upon the arrival of the SiPMs.

\vspace{-2em}

\subsection{Simulation}

Simulation studies using Geant4 have shown promising results for the HiDRa prototype. 

\begin{figure}[h]
\centering
\includegraphics[width=6.5cm,clip]{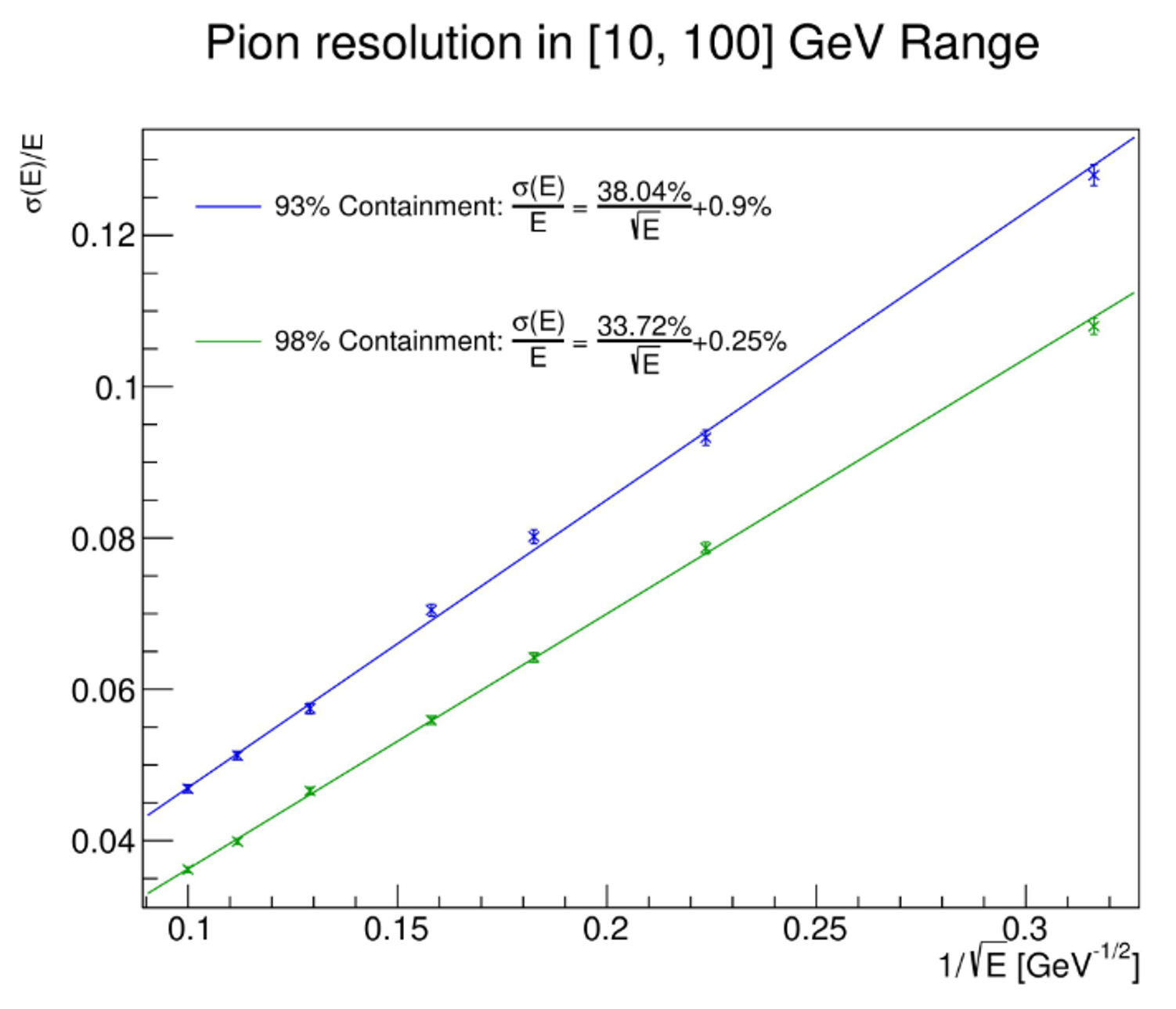}
\caption{Expected energy resolution for hadron showers from pion beams, based on a Geant4 simulation, for two calorimeter configurations: the HiDRa layout (blue points) and a larger detector containing six times more modules (green points).}
\label{fig::hidraresN}      
\end{figure}

Particle leakage for hadronic showers mainly occurs from the sides of the detector. The $65 \times 65\, \cm^2$ cross-section can contain about $93\%$ of the energy from a 40~GeV pion beam, as detailed further in \cite{PARETI}. Despite the expected leakage, good energy resolution is still achieved. This is illustrated in Figure \ref{fig::hidraresN}, which shows the resolution as a function of the $\pi^+$ energy in the range from 10 to 100~GeV. The figure also  shows that by enhancing containment to 98\% —achieved by using a detector layout with 480 modules instead of 80— the stochastic term could be reduced to approximately $33.7\% / \sqrt{E}$. This adjustment more closely aligns with the performances expected by using such dual-readout technology in a future full-coverage calorimeter.

\begin{figure}[h]
\centering
\includegraphics[width=6.55cm,clip]{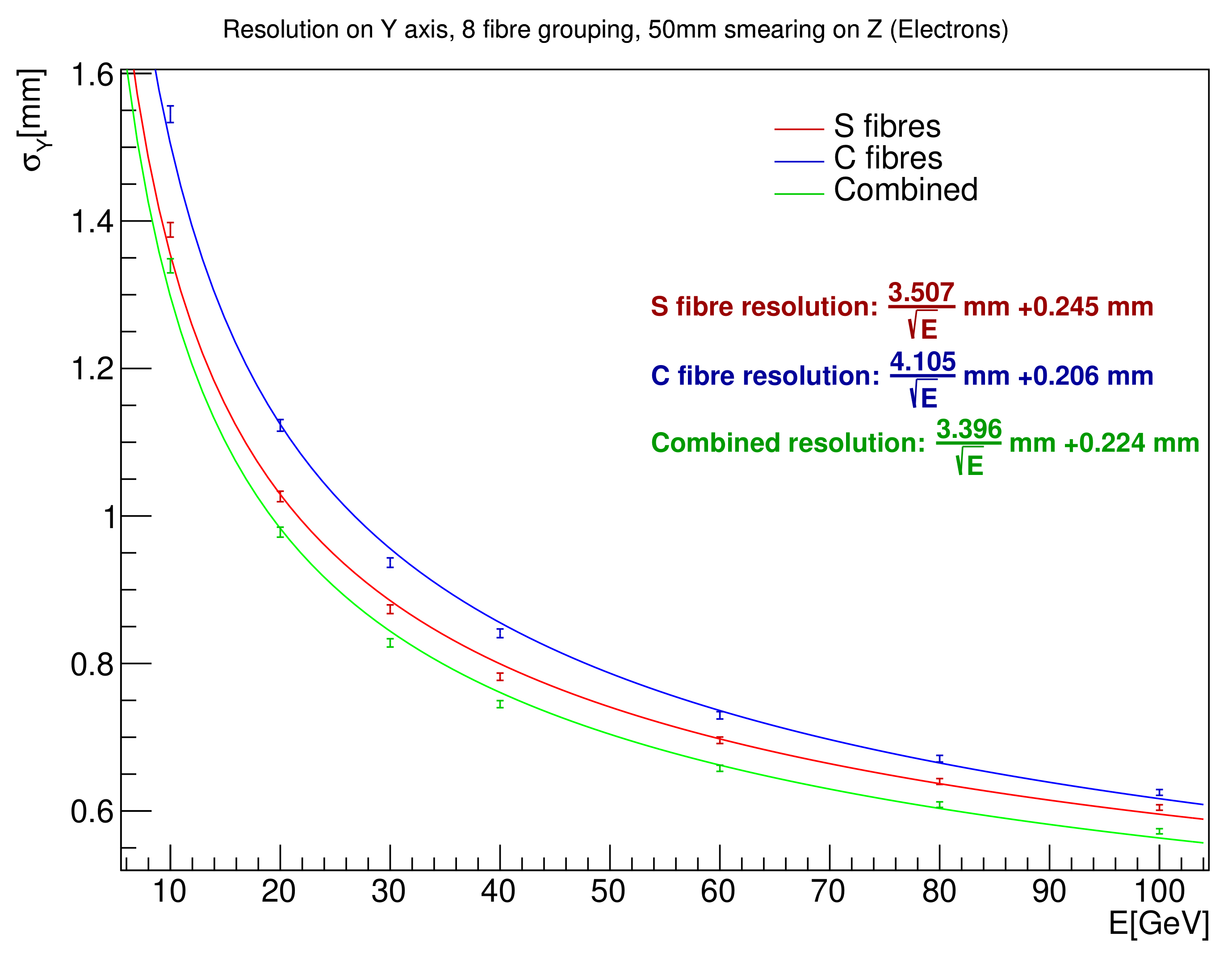}
\caption{Spatial resolution from showers induced by positrons. The 8-channels grouping foreseen by the HiDRa front-end boards design is simulated.}
\label{fig::hidraspatial}      
\end{figure}

Additional insights from the simulation include spatial resolution, which is reconstructed based on the center of gravity of the involved fibers. The results not only demonstrate millimeter-level accuracy achieved with the highly granular readout, but also confirm that the fiber grouping has a negligible effect on the spatial resolution (eight fibers extend approximately 16~mm in one direction, while the Moli\`ere radius in HiDRa is around 25~mm). The spatial resolution for electromagnetic showers from positrons, obtained by aggregating signals from eight SiPMs, is presented in Figure \ref{fig::hidraspatial}.

\section{Conclusions}

Future experiments at colliders will benefit significantly from precise hadron jet measurements. The dual-readout calorimetry technique, combined with SiPM readout, offers substantial improvements in energy resolution and opens up advanced reconstruction capabilities. This paper highlighted the construction, integration, testing and simulation phases of the existing dual-readout calorimeter based on capillary tubes and single-fiber readout. It has outlined key steps in identifying scalable and efficient assembly methods, validated on test beams and through simulations, which show promising results. This fully supports the use of the proposed technology in future calorimeters.

\end{document}